\documentclass[pra,aps,showpacs]{revtex4} 
\usepackage{epsfig} 
\begin{document} 
\title{Relation between geometric phases of entangled bi-partite  
systems and their subsystems} 
\author{D.M. Tong $^1$, E. Sj\"{o}qvist$^2$\footnote{Electronic  
address: eriks@kvac.uu.se}, L.C. Kwek$^{1,3}$, 
C.H. Oh$^1$\footnote{Electronic address: phyohch@nus.edu.sg}, and 
M. Ericsson$^{2,4}$\footnote{Electronic address: mericssn@uiuc.edu}} 
\affiliation{$^1$Department of Physics, National University of 
Singapore, 10 Kent Ridge Crescent, Singapore 119260 \\  
$^2$Department of Quantum Chemistry, Uppsala University, Box 518, 
Se-751 20 Uppsala, Sweden \\  
$^3$National Institute of Education, Nanyang Technological 
University, 1 Nanyang Walk, Singapore 639798 \\ 
$^4$Department of Physics, University of Illinois at 
Urbana-Champaign, 1110 West Green Street, Urbana, IL 61801-3080, 
USA} 
\date{\today} 
\begin{abstract} 
This paper focuses on the geometric phase of entangled states of 
bi-partite systems under bi-local unitary evolution. We investigate 
the relation between the geometric phase of the system and those of 
the subsystems.  It is shown that (1) the geometric phase of cyclic 
entangled states with non-degenerate eigenvalues can always be 
decomposed into a sum of weighted non-modular pure state phases 
pertaining to the separable components of the Schmidt decomposition, 
though the same cannot be said in the non-cyclic case, and (2) the 
geometric phase of the mixed state of one subsystem is generally 
different from that of the entangled state even by keeping the other 
subsystem fixed, but the two phases are the same when the evolution 
operator satisfies conditions where each component in the Schmidt 
decomposition is parallel transported. 
\end{abstract} 
\maketitle 
\section{Introduction} 
In 1984, Berry {\cite{Berry}} proposed in a seminal paper that 
a quantum system in a pure state undergoing adiabatic cyclic 
evolution acquires a geometric phase. This discovery has prompted 
a myriad of activities on various aspects of geometric phase in 
many areas of physics ranging from optical fibers to anyons. Simon 
\cite{Simon} subsequently recast the mathematical formalism of 
Berry's phase within the language of differential geometry and 
fiber bundles. While it is possible to consider Berry's phase 
under adiabatic evolution, the extension to non-adiabatic 
evolution is usually non-trivial. The general formalism for the 
non-adiabatic extension was formulated by Aharonov and Anandan 
\cite{Aharonov,Anandan}. Samuel and Bhandari \cite{Samuel} 
further generalized the geometric phase by extending it to 
non-cyclic evolution and sequential measurements. Further  
relaxation on the adiabatic, unitary, and cyclic properties  
of the evolution have since been carried out  
\cite{Mukunda,Pati,Manini}. 
 
The concept of geometric phase of mixed states has also been 
developed.  Uhlmann {\cite{Uhlmann}} was probably first to address 
this issue within the mathematical context of purification. 
Sj\"oqvist {\it et al.} \cite{Sjoqvistm} have introduced a new 
formalism that defines the mixed state geometric phase within the 
experimental context of quantum interferometry. As pointed out by 
Slater \cite{Slater}, these two approaches are not equivalent, and 
Ericsson {\it et al.} \cite{Ericsson} have recently shown that the 
parallel transport conditions used in the two approaches lead to 
generically distinct phase holonomy effects for entangled systems 
undergoing certain local unitary transformations. Singh {\it et al.} 
\cite{Kuldip} have given a kinematic formalism of the mixed state 
geometric phase with non-degenerate eigenvalues and generalized the 
analysis of Ref. \cite{Sjoqvistm} to degenerate states.  Extension of 
Ref. \cite{Sjoqvistm} to the off-diagonal case \cite{Filipp} and 
completely positive maps \cite{ericsson03} have also been given. An 
experimental test of Ref. \cite{Sjoqvistm} in the qubit case has been 
reported \cite{du03}, using nuclear magnetic resonance technique. 
 
The geometric phase of entangled states should be another issue worthy 
of attention. It may hold a potential application in holonomic quantum 
computation since the study of entangled spin systems effectively 
allows us to contemplate the design of a solid state quantum computer 
{\cite{burkard}}. Moreover, it was found that one could in principle 
devise more robust fault-tolerant quantum computations using the 
notion of geometric phase in designing a conditional phase shift gate 
\cite{jones,ekert2000,wang}.  Since the geometric phase depends solely 
on the geometry of the intrinsic spin space, it is deemed to be less 
susceptible to noise from the environment. As for the geometric phase 
of entangled states itself, Sj\"{o}qvist \cite{Sjoqvistp} considered 
the geometric phase for a pair of entangled spin$-\frac{1}{2}$ systems 
in a time-independent uniform magnetic field, and the relative phase 
for polarization-entangled two-photon systems has been considered by 
Hessmo and Sj\"{o}qvist \cite{Hessmo}. Tong {\it et al.} \cite{Oh} 
calculated the geometric phase for a pair of entangled 
spin$-\frac{1}{2}$ systems in a rotating magnetic field. Another 
interesting question, which has been mentioned in the previous papers 
but has not been completely discussed, is the relation between 
geometric phase of the entangled state and those of the subsystems. 
 
In this paper, we consider a general entangled bi-partite system under 
an arbitrary bi-local unitary evolution. We extend the conclusions 
obtained from the special mode in \cite{Oh} to the general case.  
That is, we prove that the geometric phase for cyclic entangled states 
with non-degenerate eigenvalues under bi-local unitary evolution can 
always be decomposed into a sum of weighted non-modular pure state 
phases pertaining to the separable components of the Schmidt 
decomposition, irrespective of forms of local evolution 
operators. However, we also see that this property does not manifest 
itself in the case of non-cyclic evolutions.  Moreover, we investigate 
the relation between the geometric phase of pure entangled states of 
the system and that of mixed states of the subsystem and conclude that 
the geometric phase of mixed states of one subsystem is different from 
that of entangled states in general even if we do not act on the other 
subsystem. We also point out that the two phases are equal when the 
evolution operator only acts on the considered subsystem and satisfies 
conditions where each pure state component of the Schmidt 
decomposition is parallel transported.

\section{Non-cyclic geometric phase of entangled state} 
We begin by considering a quantum system $S$ consisting of two 
subsystems $S_a$ and $S_b$ subject to the bi-local unitary evolution 
$U(t)=U_a(t)\otimes U_b(t)$. The states of the system belong to 
the Hilbert space ${\cal H}={\cal H}_a \otimes{\cal H}_b$, where 
${\cal H}_a$ and ${\cal H}_b$  are two complex Hilbert spaces 
with dimensions $N_a$ and $N_b$, respectively. Vectors in 
${\cal H}_a \otimes {\cal H}_b$ can be expanded as Schmidt 
decompositions. Thus, any normalized initial state of the system 
can be written as 
\begin{eqnarray} 
|\Psi (0)\rangle = 
\sum \limits_{i=1}^{N} \sqrt{p_i} |\mu_i\rangle 
\otimes |\nu_i\rangle, 
\label{Psi} 
\end{eqnarray} 
where $|\mu_i\rangle $ and $|\nu_i\rangle$ are orthonormal 
bases of ${\cal H}_a$ and ${\cal H}_b$, respectively, 
$N = {\text{min}}\{N_a,~N_b\}$, and the Schmidt coefficients 
$\{ p_i \}$ fulfill $p_1 + \ldots + p_N =1$. When 
$U_a(t) \otimes U_b(t)$ acts on $|\Psi (0) \rangle$, we obtain 
\begin{eqnarray} 
|\Psi (t)\rangle =\sum\limits_{i=1}^{N} \sqrt{p_i} |\mu'_i(t) 
\rangle \otimes |\nu'_i(t)\rangle, 
\label{Psit} 
\end{eqnarray} 
where $|\mu'_i(t)\rangle= U_a(t) |\mu_i\rangle$ and $|\nu'_i(t) 
\rangle = U_b(t) |\nu_i\rangle $.  
 
Since $|\Psi (t) \rangle$ is a pure state, its geometric phase can be 
obtained by removing the dynamical phase from the total phase. As we 
know, when a pure state evolves from $t=0$ to $t=\tau$ along a path 
$C$~in projective Hilbert space, the non-adiabatic geometric phase can 
be obtained as $\gamma(\tau)=\alpha(\tau)-\beta(\tau)$ with total 
phase $ \alpha(\tau)=\arg\langle \Psi (0)|\Psi (\tau) 
\rangle$ and dynamical phase $\beta(\tau)=- i\int_0^\tau 
\langle \Psi (t) |\dot \Psi (t)\rangle dt$. Hereafter, we use 
$\alpha$, $\beta$, and $\gamma$ to mark total, dynamical, and 
geometric phases, respectively; and we use $t$, $\tau$, and $T$ to 
represent instantaneous time, finite time, and period of cyclic 
evolution, respectively. With Eqs.~(\ref{Psi}) and (\ref{Psit}), 
we obtain the total, dynamical, and geometric phase of the entangled 
state $|\Psi (t)\rangle$ under a bi-local unitary evolution 
$U_a(t)\otimes U_b(t)$ as 
\begin{eqnarray} 
\alpha_{ab}(\tau) & = & \arg \sum\limits_{i,j=1}^{N}  
\sqrt{p_ip_j} \langle \mu_i | U_a (\tau) |\mu_j\rangle  
\langle \nu_i |U_b(\tau)| \nu_j \rangle,  
\label{aab} \\ 
 & &  
\nonumber \\ 
\beta_{ab}(\tau) & = & \sum\limits_{i=1}^{N}  
p_i \left( - i\int_0^\tau \langle \mu_i |U_a^{\dagger} (t)  
\dot U_a(t) |\mu_i\rangle dt \right)  
\nonumber \\ 
 & & +\sum\limits_{i=1}^{N} p_i \left( - i\int_0^\tau \langle \nu_i 
|U_b^{\dagger} (t)\dot U_b(t) |\nu_i\rangle dt \right) ,  
\label{bab}\\ 
 & &  
\nonumber \\  
\gamma_{ab}(\tau) & = & \alpha_{ab}(\tau)-\beta_{ab}(\tau). 
\label{gab} 
\end{eqnarray}  
Eq.~(\ref{bab}) entails that the dynamical phase can always be 
separated into two parts corresponding to the evolution of each 
of the subsystems $S_a$ and $S_b$. However, the total phase as well 
as the geometric phase cannot be separated into two parts in general. 
The latter observation arises primarily from the entanglement of the 
two subsystems. 
 
\section{Cyclic geometric phase of entangled states} 
In this section, we specialize the above discussion to cyclic states 
with non-degenerate Schmidt coefficients $p_i$. Such states are 
characterized by the existence of a period $T$ such that 
\begin{eqnarray} 
|\Psi (T)\rangle = e^{i\alpha_{ab}(T)} |\Psi (0)\rangle , 
\label{Psic} 
\end{eqnarray} 
that is, 
\begin{equation} 
\sum\limits_{i=1}^{N} \sqrt{p_i} |\mu'_i(T)\rangle \otimes | 
\nu'_i(T)\rangle = e^{i\alpha_{ab}(T)} \sum\limits_{i=1}^{N} 
\sqrt{p_i} |\mu_i\rangle \otimes |\nu_i\rangle . 
\label{Psit=Psi0} 
\end{equation} 
As $U_a(T)$ and $U_b(T)$ are unitary, the vectors $|\mu'_i(T) 
\rangle$ and $|\nu'_i(T)\rangle$ are also orthonormal bases  
of ${\cal H}_a$ and ${\cal H}_b$, respectively. Moreover,  
under consideration that the $p_i$'s are non-degenerate, i.e.,  
that $p_i \neq p_j$ for all pairs $i,j \neq i$, the 
Schmidt decomposition is unique. So, we have $|\mu'_i(T) \rangle 
\otimes |\nu'_i(T)\rangle = e^{i\alpha_{ab}(T)} |\mu_i 
\rangle \otimes |\nu_i\rangle$, which implies 
\begin{eqnarray} 
\alpha_{ab}(T) = \arg \big[ \langle\mu_i |U_a(T)|\mu_i\rangle \langle 
\nu_i |U_b(T)| \nu_i\rangle \big] , 
\label{ab=ab} 
\end{eqnarray} 
where $\arg \in [0,2\pi)$. Since the left-hand side of 
Eq.~(\ref{ab=ab}) is independent of the summation index 
$i$, $\alpha_{ab}(T)$ can be written as 
\begin{eqnarray} 
\alpha_{ab}(T) & = & \sum\limits_{i=1}^{N} 
p_i \arg \big[ \langle\mu_i |U_a(T) |\mu_i\rangle 
\langle \nu_i |U_b(T)|\nu_i \rangle \big] 
\nonumber\\ 
 & & = \sum\limits_{i=1}^{N} p_i \big[ \arg \langle\mu_i | 
U_a(T) |\mu_i\rangle + \arg \langle \nu_i |U_b(T)|  
\nu_i\rangle +2\pi \tilde{n}_i \big]  
\label{ab=sab} 
\end{eqnarray} 
where $\tilde{n}_{i}$ are integers chosen as  
\begin{eqnarray}  
\tilde{n}_i & = & 0, \ \ {\textrm{if}} \ \arg \langle\mu_i | 
U_a(T) |\mu_i\rangle + \arg \langle \nu_i |U_b(T)|  
\nu_i\rangle < 2\pi ,  
\nonumber \\   
\tilde{n}_i & = & -1, \ \ {\textrm{if}} \ \arg \langle\mu_i | 
U_a(T) |\mu_i\rangle + \arg \langle \nu_i |U_b(T)|  
\nu_i\rangle \geq 2\pi .  
\end{eqnarray} 
Substituting Eqs.~(\ref{bab}) and (\ref{ab=sab}) into 
Eq.~(\ref{gab}) yields 
\begin{eqnarray} 
\gamma_{ab}(T) = \sum\limits_{i=1}^N p_i \big[ \gamma_{ai}(T) +  
\gamma_{bi}(T) + 2\pi n_{i} \big] ,  
\label{a=a} 
\end{eqnarray} 
where 
\begin{eqnarray} 
\gamma_{ai}(T)&=& \arg \left( \langle\mu_i |U_a(T)| \mu_i\rangle e^{- 
\int_0^T \langle \mu_i |U_a^{\dagger} (t) \dot U_a(t) 
|\mu_i\rangle dt } \right) , 
\nonumber \\ 
\gamma_{bi}(T) & = & \arg \left( \langle \nu_i |U_b(T) 
|\nu_i\rangle e^{-\int_0^T \langle \nu_i | 
U_b^{\dagger} (t) \dot U_b(t) | \nu_i\rangle dt } \right) , 
\label{gabi} 
\end{eqnarray} 
are just the $2\pi-$modular geometric phases of the pure states 
$|\mu_i\rangle$ and $|\nu_i\rangle$, respectively. The ``winding 
numbers'' $n_i$ are integers and originate from the non-modular 
nature of the pure state dynamical phases \cite{Mukunda} and the 
integers $\tilde{n}_i$.  Eq.~(\ref{a=a}) shows that the cyclic 
geometric phase for non-degenerate entangled states under a bi-local 
unitary evolution can always be decomposed into a sum of weighted pure 
state phases pertaining to the evolution of each Schmidt 
component. This is primarily a result of the uniqueness of the Schmidt 
decomposition, which entails that the Schmidt basis of the initial and 
final state must be identical if all $p_i$ are different. 
 
The contribution from the non-modular nature of the dynamical phases 
to the winding numbers $n_i$ can be determined in a history-dependent 
manner \cite{bhandari91,bhandari02} by continuously monitoring the 
dynamical phase for each separate component of the Schmidt 
decomposition on the time interval $[0,T]$. In such a procedure, each 
$2\pi n_i$ is the modulus $2\pi$ remainder of the corresponding 
dynamical phase. 
 
To illustrate the significance of the winding numbers, consider  
a pair of qubits (two-level systems) being initially in the  
entangled state 
\begin{equation} 
|\Psi (0) \rangle = \sqrt{p_0} |0_a \rangle 
\otimes |0_b \rangle + \sqrt{p_1} |1_a \rangle 
\otimes |1_b \rangle 
\end{equation} 
with $p_0 \neq p_1$, and evolving under influence of the  
time-independent Hamiltonian 
\begin{equation} 
H = \hbar \omega |1_a \rangle \langle 1_a| \otimes I_b 
\end{equation} 
with $\omega>0$ and $I_b$ the identity operator on  
${\cal H}_b$. As the Schmidt components $|0_a \rangle 
\otimes |0_b \rangle$ and $|1_a \rangle \otimes |1_b \rangle$ 
are eigenstates of $H$, their corresponding $2\pi-$modular  
geometric phases $\{ \gamma_{a0} , \gamma_{b0} \}$ and  
$\{ \gamma_{a1} , \gamma_{b1} \}$ all vanish. However, they acquire  
different dynamical phases causing a non-trivial evolution  
of the entangled state. For one cycle $T=2\pi /\omega$ one  
obtains $n_0 = 0$ and $n_1=1$, yielding the total geometric phase 
\begin{equation} 
\gamma_{ab} (T) = 2\pi p_1 , 
\end{equation} 
which is non-trivial for entangled states. 
 
\section{Comparing the phase of entangled states with that  
of mixed states}  
The above result concerning the phase of cyclic entangled states may 
provide some insight regarding the geometric phase of mixed states. 
Certainly, the state given by Eq.~(\ref{Psit}) is a pure state with 
density matrix $\rho_{ab}(t)= |\Psi (t)\rangle \langle \Psi (t) |$. 
However, if we trace out the state of subsystem $S_b$, we obtain the 
reduced density matrix $\rho_a$ corresponding to mixed states of 
subsystem $S_a$. By the reduced density matrix, we can deduce the 
geometric phase $\gamma_a^M(T)$ of a mixed state. Here, we wish to 
examine, in the case where $U_b(t)$ is the identity map $I_b$, the  
relation between $\gamma_{ab} (T)$ and $\gamma_a^M (T)$. 
 
From Eqs.~(\ref{Psi}) and (\ref{Psit}), tracing out the state of 
subsystem $S_b$, we obtain the evolution of the reduced density matrix 
for subsystem $S_a$ as 
\begin{eqnarray} 
\rho_{a}(t)=U_a(t) \rho_a(0) U_a^{\dagger} (t) 
\label{rho} 
\end{eqnarray} 
with 
\begin{eqnarray} 
\rho_{a}(0)=\sum\limits_{i=1}^{N} p_i |\mu_i \rangle \langle 
\mu_i | . 
\label{rho0} 
\end{eqnarray} 
For non-degenerate $p_i$, the geometric phase of $\rho_{a}(t)$ is 
found as \cite{Sjoqvistm,Kuldip} 
\begin{equation} 
\gamma_a^M(\tau) =  
\arg \left( \sum\limits_{i=1}^{N} p_i\langle \mu_i | 
U_a(\tau)| \mu_i\rangle e^{-\int_0^\tau \langle \mu_i | 
U^{\dagger}_a(t)\dot U(t)| \mu_i\rangle dt} \right) . 
\label{mg} 
\end{equation} 
This expression is valid for both cyclic and non-cyclic states. 
In the cyclic case $\tau=T$ we have $|\langle \mu_i | 
U_a(T)| \mu_i\rangle | = 1$, and the above equation can be 
written as 
\begin{eqnarray} 
\gamma_a^M (T) = \arg \left( \sum\limits_{i=1}^{N} p_i 
e^{i\gamma_{ai}(T)} \right) , 
\label{mgl} 
\end{eqnarray} 
where $ \gamma_{ai}(T) $ is the geometric phase of the pure state 
$|\mu_i\rangle$. Comparing Eq.~(\ref{mgl}) with Eq.~(\ref{a=a}) for  
$U_b (t) = I_b$, we find $\gamma_a^M (T)\neq \gamma_{ab} (T)$.  
Thus, the cyclic geometric phase of the whole system is in general  
different from that of the mixed state of the considered subsystem,  
basically because the former is a weighted sum of pure state   
phases while the latter is a weighted sum of pure state phase  
factors.  
 
Similarly, for non-cyclic evolution with $U(t)=U_a(t)\otimes I_b$,   
we have in general  
\begin{eqnarray} 
\gamma_{ab}(\tau) \equiv \gamma_a(\tau) = \arg \sum\limits_{i=1}^{N} 
p_i\langle \mu_i |U_a(\tau)|\mu_i\rangle + 
i\int_0^\tau\sum\limits_{i=1}^{N} p_i  \langle \mu_i | 
U_a^{\dagger} (t)\dot U_a(t) | \mu_i\rangle dt   
\label{Ga} 
\end{eqnarray}  
and  
\begin{eqnarray} 
\gamma_a^M (\tau) = \arg \left( \sum\limits_{i=1}^{N} p_i|\langle \mu_i | 
U_a(\tau)| \mu_i\rangle | 
e^{i\gamma_{ai}(\tau)} \right) .  
\end{eqnarray}  
Thus, the geometric phase of the system $S$ is dependent upon 
$|\mu_i\rangle$ of subsystem $S_a$ but independent of $|\nu_i\rangle$ 
of $S_b$. Only the evolution of subsystem $S_a$ contributes to the 
geometric phase of the pure state system. Yet, we find that, even in 
the case when the system's geometric phase is completely determined by 
the evolution of subsystem $S_a$ and the Schmidt coefficients 
$\{p_i\}$, $\gamma_a(\tau) $ is generally different from 
$\gamma_a^M(\tau)$ both in the cyclic and non-cyclic case. It may  
seem unexpected, because only subsystem $S_a$ experiences a unitary 
evolution while subsystem $S_b$ is unaffected. The geometric phase of 
the system $S$ is attributed to $U_a(t)$ only, and it seems natural to 
expect the phase obtained by the system to be same as that obtained by 
subsystem $S_a$, while regarding it as mixed state. However, they are 
different. This shows that the geometric phase of an entangled 
bi-partite system is always affected by both subsystems. 
 
\section{Phase relations under parallel transport conditions} 
In this section, we again restrict our discussion to the case where 
$U(t)=U_a(t)\otimes I_b$. As pointed out above, even in this case, the 
geometric phase of the system, which is determined only by the 
evolution of subsystem $S_a$ and the Schmidt coefficients $\{p_i\}$, 
is generally different from that of the corresponding mixed state. We 
now try to find the reason for the difference and give conditions 
under which the two phases are equal. 
 
Under the evolution $U(t)=U_a(t)\otimes I_b$, the  state of the whole 
system is $|\Psi(t) \rangle=\sum\limits_{i=1}^{N}\sqrt{p_i}\quad 
[U_a(t)|\mu_i\rangle] \otimes |\nu_i\rangle$, which can also be 
expressed as the density matrix  
\begin{eqnarray} 
\rho(t) = \sum\limits_{i=1}^N \sqrt{p_ip_j} [U_a(t) |\mu_i \rangle 
\langle \mu_j |U_a^{\dagger} (t)] \otimes 
|\nu_i \rangle \langle \nu_j | . 
\label{rhoab} 
\end{eqnarray} 
The corresponding mixed state of the subsystem $S_a$ is 
\begin{eqnarray} 
\rho_{a}(t)={\text{Tr}}_b \rho(t)= 
\sum\limits_{i=1}^{N} p_i U_a(t) |\mu_i \rangle \langle 
\mu_i | U_a^{\dagger} (t). 
\label{rho2} 
\end{eqnarray} 
When $U_a(t)$ is given, the states $\rho(t)$ and $\rho_a(t)$ are 
definite, but when $\rho(t)$ or $\rho_a(t)$ is given, the evolution 
operator is not unique. That is, for a given path in state 
space, there are infinitely many unitary operators that realize the 
same path and so give the same geometric phase. All the operators form 
an equivalence set: two evolution operators are `equivalent' if  
and only if they realize the same path. If we know any one operator  
out of the equivalence set, say $\tilde{U}_a(t) $, we can   
write down all the operators of the set. For $\rho(t)$, the  
equivalence set is 
\begin{equation} 
{\cal{S}}_1 = \left\{ \tilde {U}_a(t)e^{i\theta (t)} \right\},  
\end{equation} 
where $\theta (t)$ is an arbitrary real-valued gauge function of $t$ 
with $\theta (0)=0$. For state $\rho_a(t)$, the equivalence set is 
\begin{equation} 
{\cal{S}}_2 = \left\{ \tilde{U}_a(t)\sum\limits_{i=1}^{N}  
e^{i\theta_i(t)} |\mu_i \rangle \langle \mu_i | \right\} , 
\end{equation}  
where $\theta_i(t)$, $i=1,2,...,N$, are arbitrary real-valued gauge 
functions of $t$ with $\theta_i (0)=0$. We see that the two sets are 
different in general and ${\cal{S}}_1 \subset {\cal{S}}_2$, which 
shows that the evolution operators that give the same path for 
$\rho_a(t)$ may give different paths for $\rho(t)$. So the two kinds 
of geometric phases $\gamma_a(\tau)$ and $\gamma_a^M(\tau)$ cannot be 
the same in general, otherwise they should be associated with the same 
equivalence sets of evolution operators. To see exactly the difference 
between the two phases, we substitute 
\begin{eqnarray} 
U_a(t) = \tilde{U}_a (t) \sum\limits_{i=1}^{N} 
e^{i\theta_i(t)} |\mu_i \rangle \langle \mu_i | 
\label{Uae} 
\end{eqnarray} 
into Eqs.~(\ref{mg}) and (\ref{Ga}), respectively, and get 
\begin{eqnarray} 
\gamma_a^M(\tau) & = & 
\arg \left( \sum\limits_{i=1}^{N} p_i\langle \mu_i | 
\tilde{U}_a(\tau) |\mu_i\rangle e^{-\int_0^\tau \langle \mu_i | 
\tilde{U}^{\dagger}_a(t)\dot{\tilde{U}}_a(t)|\mu_i\rangle dt}  
\right) ,  
\label{mg3} 
\\ 
\gamma_a(\tau) & = & \arg\sum\limits_{i=1}^{N} p_i\langle \mu_i | 
\tilde{U}_a (\tau)|\mu_i\rangle e^{i\theta_i(t)} 
\nonumber \\ 
 & & + i\int_0^\tau \sum\limits_{i=1}^{N} p_i \langle \mu_i | 
\tilde{U}_a^{\dagger} (t)\dot{\tilde{U}}_a(t) | \mu_i\rangle dt - 
\sum\limits_{i=1}^{N} p_i\theta_i(\tau). 
\label{Ga3} 
\end{eqnarray} 
We see that $\gamma_a^M(\tau)$ is invariant under choice of member  
in ${\cal S}_2$, while $\gamma_a(\tau)$ is not as it depends upon  
$\theta_i(t)$. 
 
With the above analysis, we can conclude that, for a given local 
evolution operator $U_a(t)$, the two phases $\gamma_a(\tau)$ and 
$\gamma_a^M(\tau)$ are different in general.  But when can they 
be the same, that is, for what kinds of $U_a(t)$ can we consider 
the two phases to be the same?  We prove that when the evolution 
operator $U_a(t)$ satisfies `the stronger parallel transport 
conditions' \cite{Sjoqvistm} 
\begin{eqnarray} 
\langle \mu_i |U_a^{\dagger} (t)\dot U_a(t) 
|\mu_i\rangle=0,~~~~i=1,2,...N, 
\label{spc} 
\end{eqnarray} 
the two phases are the same. Substituting Eq.~(\ref{spc}) into 
Eqs.~(\ref{mg}) and (\ref{Ga}), we find 
\begin{eqnarray} 
\gamma_a(\tau) = \gamma_a^M(\tau) =  
\arg \left( \sum\limits_{i=1}^{N} p_i\langle 
\mu_i |U_a(\tau)|\mu_i\rangle \right) ,  
\label{a=am} 
\end{eqnarray} 
and 
\begin{eqnarray} 
\gamma_b(\tau) =\gamma_b^M(\tau)=0. 
\label{b=bm} 
\end{eqnarray} 
Eqs.~(\ref{a=am}) and (\ref{b=bm}) show that the two kinds of 
geometric phase for unitarities of the form $U_a (t) \otimes I_b$ 
are always the same when the evolution operator $U_a(t)$ satisfies 
the stronger parallel transport conditions Eq. (\ref{spc}). 
 
Substituting $U_a(t) = V_a(t) \sum\limits_{i=1}^{N} e^{i\theta_i(t)}  
|\mu_i \rangle \langle \mu_i |$ into Eq.~(\ref{spc}), we get the 
general form of evolution operators satisfying the stronger 
parallel transport conditions  
\begin{eqnarray} 
U_a(t) =V_a(t)\sum\limits_{i=1}^N |\mu_i\rangle\langle\mu_i | 
e^{-\int_0^t \langle \mu_i | V_a^{\dagger} (t) \dot V_a (t) 
|\mu_i\rangle dt}, 
\label{sua} 
\end{eqnarray} 
where $V_a(t)$ is an arbitrary unitary operator. So we see that when 
$U_a(t)$ holds the form of Eq.~(\ref{sua}), the geometric phase of the 
pure state of the system under the evolution $U(t)=U_a(t)\otimes I_b$ is 
the same as that of the mixed state of subsystem $S_a$ under the 
evolution $U_a(t)$. The phase relations are shown as Eqs.~(\ref{a=am}) 
and (\ref{b=bm}). 

\begin{figure}[ht!] 
\begin{center} 
\includegraphics[width=8 cm]{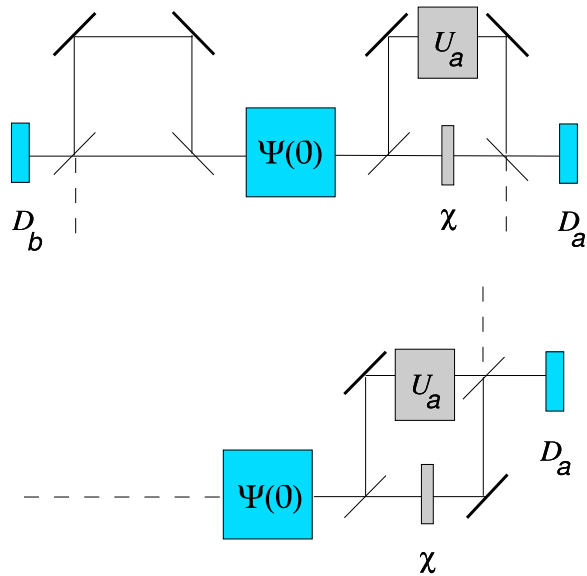} 
\end{center} 
\caption{Two-particle interferometry setups to measure phases 
for internal degrees of freedom in entangled pure state 
$|\Psi (0)\rangle$. In the upper panel, the pure state phase 
$\arg \langle \Psi (0)| U_a (\tau) \otimes I_b | \Psi (0) \rangle$ 
is measured as a shift in the coincidence interference oscillations 
obtained at detectors $D_a$ and $D_b$ by varying the $U(1)$ phase 
$\chi$. In the lower panel, the mixed state phase $\arg{\text{Tr}} 
\big( \rho_a (0) U_a (\tau) \big)$, with $\rho_a (0) = 
{\text{Tr}}_b |\Psi (0)\rangle \langle \Psi (0)|$ the mixed input 
state of subsystem $a$, is similarly measured as a shift at $D_a$ 
by ignoring the subsystem $b$.} 
\end{figure} 
 
It might also be useful to consider the problem from an operational
point of view. Let us first consider the two-particle Franson type 
interferometry setup shown in the upper panel of Fig. 1, 
with internal pure state $|\Psi (0) \rangle$ affected by 
$U_a (t) \otimes I_b$ in the longer arms and by a $U(1)$ shift 
$\chi$ to one of the shorter arms. To observe the relative phase 
between $|\Psi (0) \rangle$ and $|\Psi (\tau) \rangle = U_a (\tau) 
\otimes I_b |\Psi (0) \rangle$, we require that the source produces 
particle pairs randomly. Then, if the particles arrive at the two 
detectors $D_a$ and $D_b$ simultaneously, they both either took the 
upper path or the lower path, so that the intensity detected in 
coincidence at $D_a$ and $D_b$ becomes \cite{Hessmo} 
\begin{eqnarray} 
{\cal I}_{ab} & \propto &  
\Big| e^{i\chi} |\Psi (0) \rangle + |\Psi (\tau) \rangle \Big|^2 
\nonumber \\ 
 & = & 2 + 2 \Big| \langle \Psi (0) | U_a (\tau) \otimes I_b 
|\Psi (0) \rangle \Big| \cos \Big[ \chi - \arg  \langle \Psi (0) 
|U_a (\tau) \otimes I_b |\Psi (0) \rangle \Big] . 
\end{eqnarray}
Thus, the relative phase $\arg  \langle \Psi (0) |U_a (\tau) 
\otimes I_b |\Psi (0) \rangle$ shifts the interference oscillations 
when $\chi$ varies. To observe the phase of $\rho_a (t) = 
{\text{Tr}}_b |\Psi (t) \rangle \langle \Psi (t)|$ acquired 
by subsystem $a$ alone, we instead use the setup shown in the lower 
panel of Fig. 1 with the same source but where subsystem $b$ 
is ignored. Here, the output intensity at $D_a$ reads \cite{Sjoqvistm} 
\begin{equation} 
{\cal I}_a \propto 2 + 2 \Big| {\text{Tr}} \big( \rho_a (0) U_a (\tau) 
\big) \Big| \cos \Big[ \chi - \arg {\text{Tr}} 
\big( \rho_a (0) U_a (\tau) \big) \Big] . 
\end{equation}
Note that ${\text{Tr}} \big( \rho_a (0) U_a (\tau) \big) = 
\langle \Psi (0) |U_a (\tau) \otimes I_b |\Psi (0) \rangle$ 
due to the triviality of the evolution of the $b$ subsystem. 

Suppose now we first arrange these experiments so that $U_a(t)$  
yields the pure state parallel transport condition $\langle 
\Psi (0)|U_a^{\dagger} (t) \dot{U}_a (t) \otimes I_b |\Psi (0) 
\rangle = 0$. In this case, the interference pattern obtained by 
measuring on both subsystems in coincidence would be shifted by 
the pure state geometric phase $\gamma_a (\tau)$. On other hand, 
measuring only on subsystem $a$ would yield the same shift, but 
the interpretation is different: it is the total phase 
$\alpha_a^M (\tau)$ acquired by $\rho_a (\tau)$, which is in 
general at variance with $\gamma_a^M (\tau)$ as this latter phase 
is based upon the stronger parallel transport conditions. The 
situation is different when setting up $U_a(t)$ so as to parallel 
transport $\rho_a(t)$, i.e., by implementing $U_a(t)$ according 
to Eq. (\ref{sua}). Here, the coincidence and marginal interference 
pattern are shifted by $\gamma_a (\tau)$ and $\gamma_a^M (\tau)$, 
respectively, in accordance with the above analysis.   
 
\section{Conclusion and Remarks} 
We have discussed geometric phases of entangled states of 
bi-partite systems under bi-local unitary evolution and of  
the mixed states of their subsystems. We conclude: 
 
\begin{enumerate} 
 
\item The cyclic geometric phase for entangled states with 
non-degenerate eigenvalues under bi-local unitary evolution can always 
be decomposed into a sum of weighted non-modular pure state phases 
pertaining to the separable components of the Schmidt decomposition, 
irrespective of forms of local evolution operators, though the same 
cannot be said for the non-cyclic geometric phase. 
 
\item The mixed state geometric phase of one subsystem is generally 
different from that of the entangled state even by keeping the 
other subsystem fixed, though it seems as if the two phases 
might be same. However, when the evolution operator satisfies 
the stronger parallel transport conditions for mixed states, 
the two phases are the same, and the general form of the operators 
are given. 
 
\end{enumerate} 
 
The difference between geometric phases of bi-partite systems and 
their parts has its primary cause in entanglement and thus vanishes 
in the limit of separable states. We hope that the present analysis 
may trigger multi-particle experiments to test the difference 
between phases of entangled systems and their subsystems. 
 
\section*{Acknowledgments} 
The work by Tong was supported by NUS Research Grant No. 
R-144-000-054-112. This work is also supported in part by  
the Agency for Science, Technology and Research, Singapore  
under the ASTAR Grant No. 012-104-0040 (WBS: R-144-000-071-305).  
E.S. acknowledges financial support from the Swedish Research  
Council. M.E. acknowledge financial support from the Foundation  
BLANCEFLOR Boncompagni-Ludovisi, n\'{e}e Bildt. 
 
\end{document}